\documentclass[aps,prd,groupedaddress,showpacs,nofootinbib,amssymb,balancelastpage,preprintnumbers]{revtex4}

\usepackage{graphicx,bm,color}
\usepackage{amsmath}
\usepackage{amssymb}
\usepackage{amsfonts}
\usepackage{cases}
\usepackage{cancel}
\usepackage{hyperref}

\usepackage[normalem]{ulem} 
\usepackage{color}
\usepackage{cancel}

\newcommand{\chushi}[1]{}

\begin{document}
\preprint{KEK-TH-2226}
\title{New perspective in searching for axion-like particles from flavor physics}

\author{Hiroyuki Ishida}\thanks{{\tt ishidah@post.kek.jp}}
	\affiliation{KEK Theory Center, IPNS, Tsukuba, Ibaraki 305-0801, Japan}

\author{Shinya Matsuzaki}\thanks{{\tt synya@jlu.edu.cn}}
\affiliation{Center for Theoretical Physics and College of Physics, Jilin University, Changchun, 130012,
China}

\author{Yoshihiro Shigekami}\thanks{{\tt sigekami@eken.phys.nagoya-u.ac.jp}}
\affiliation{School of Physics, Huazhong University of Science and Technology, Wuhan 430074, China}

\begin{abstract}

We propose new perspective in searching for axion-like particles (ALPs) from quark and lepton flavor physics: 
measurements of the time-dependent CP asymmetry in $B^0 \to K_S^0 \pi^0 \gamma$ and the branching ratio of $B_s \to e^\pm \mu^\mp$ decay possess, along with the anomalous magnetic moment of muon. 
In the mass range of sub-GeV, accessible by the flavorful ALPs search, the experimental sensitivity for these flavor observables reaches the maximum at around the pion mass scale (called the {\it sweetest} spots), where a couple of loopholes (unexplored regions) for the ALP parameter space have heretofore been present, because of an unavoidable contamination with pion background events. 
The proposed complementary probes can precisely determine the ALP coupling to photon at these {\it sweetest} spots/loopholes, and will significantly help cover whole parameter spaces in the ALP search including the present loopholes in the future.

\end{abstract}
\maketitle

\section{Introduction}

Axion is one of the mostly anticipated hypothetical particles predicted beyond the 
standard model (SM) of particle physics. 
The axion search has extensively been worked out with its theoretical and phenomenological interests for a long time. 
One of the strongest motivations to introduce axion is a solution for the strong CP problem~\cite{Peccei:1977hh,Peccei:1977ur,Weinberg:1977ma,Wilczek:1977pj} (the axion is called QCD axion). 

Apart from the QCD axion, axion-like pseudoscalars can also be predicted from 
a fundamental theory, which would be for instance the string theory or some unified theory. 
They are called axion-like particles (ALPs), having phenomenological properties similar to the QCD axion, such as the predominant coupling to diphoton and being long-lived. 
Though ALPs may or may not solve the strong CP problem, they can leave phenomenological footprints in high energy experiments and also astrophysical observations. 

The bulk of the ALP search has recently be extended involving flavor physics. 
Light ALPs with the mass of sub-GeV scale can have a significant impact on flavor observables, such as rare decays of heavy hadrons and leptons involving violation of the lepton flavor universality~\cite{Chang:2000ii,Izaguirre:2016dfi,Marciano:2016yhf,Bjorkeroth:2018dzu,Gavela:2019wzg,MartinCamalich:2020dfe}. 
It is anticipated to find out ALP parameter spaces in which the ALP can be dictated in both quark and lepton flavor physics. 

Of particular interest is that the flavorful ALP with such a mass can also account for the presently reported anomaly on the anomalous magnetic moment of muon, $(g-2)_\mu$~\cite{Bauer:2019gfk}. 
However, it has been shown very recently~\cite{Endo:2020mev} that when the ALP is responsible for the $(g-2)_\mu$ simply by the presence of its flavor-off diagonal couplings, the favored parameter space is excluded for the ALP mass below the sub-GeV scale by the muonium-antimuonium oscillation experiment. 
Thus the way out for the ALP to survive in the anomaly of the $(g-2)_\mu$ might be to invoke the presence of a large ALP coupling to diphoton~\cite{Bauer:2019gfk}. 

However, the case is not so optimistic: 
direct productions of the ALP through the Primakoff process by the diphoton coupling have severely been constrained by the beam dump experiments~\cite{Beacham:2019nyx}. 
Thus in a wide range of the mass below $\sim $ GeV, the stringent bounds on the ALP couplings have been placed including the diphoton coupling, and hence seems to have almost excluded the possibility for the ALP to be seen in the $(g-2)_\mu$ over the SM contributions. 

One notices, however, that there are remarkable ``loopholes" when the mass of ALPs sits around the pion mass ($\sim$ 135 -- 140 MeV). 
In this case, signals from ALPs decaying to diphoton can be hidden behind the huge background from neutral pions decaying to diphoton at beam dump experiments~\cite{Kitahara:2019lws}. 

The limit on such 140 MeV ALPs can be placed by mono-photon and tri-photon searches accessible at LEP, Tevatron and LHC, through $Z (\gamma^*) \to a + \gamma$ ($a=$ ALP)~\cite{Acciarri:1995gy,Abbiendi:2002je,Aaltonen:2013mfa,Jaeckel:2015jla,Bauer:2017ris}. 
However, this channel also contaminates with the neutral pion events such as $Z \to \pi^0 + \gamma$, which is a rare decay channel and has not observed or yet uncovered even for the SM prediction. 
A conservative bound has been addressed by simply applying the upper limit on the SM branching ratio of $Z \to \pi^0 + \gamma$ to the 140 MeV ALP~\cite{Jaeckel:2015jla,Bauer:2017ris}, which will however be insufficient 
because of the pion contamination. 
Thus, this kind of direct ALP productions in collider experiments will not supply ``clean/disentangled" signals for the 140 MeV ALP. 

\begin{figure}[t]
\centering
\includegraphics[width=0.45\textwidth,bb= 0 0 450 438]{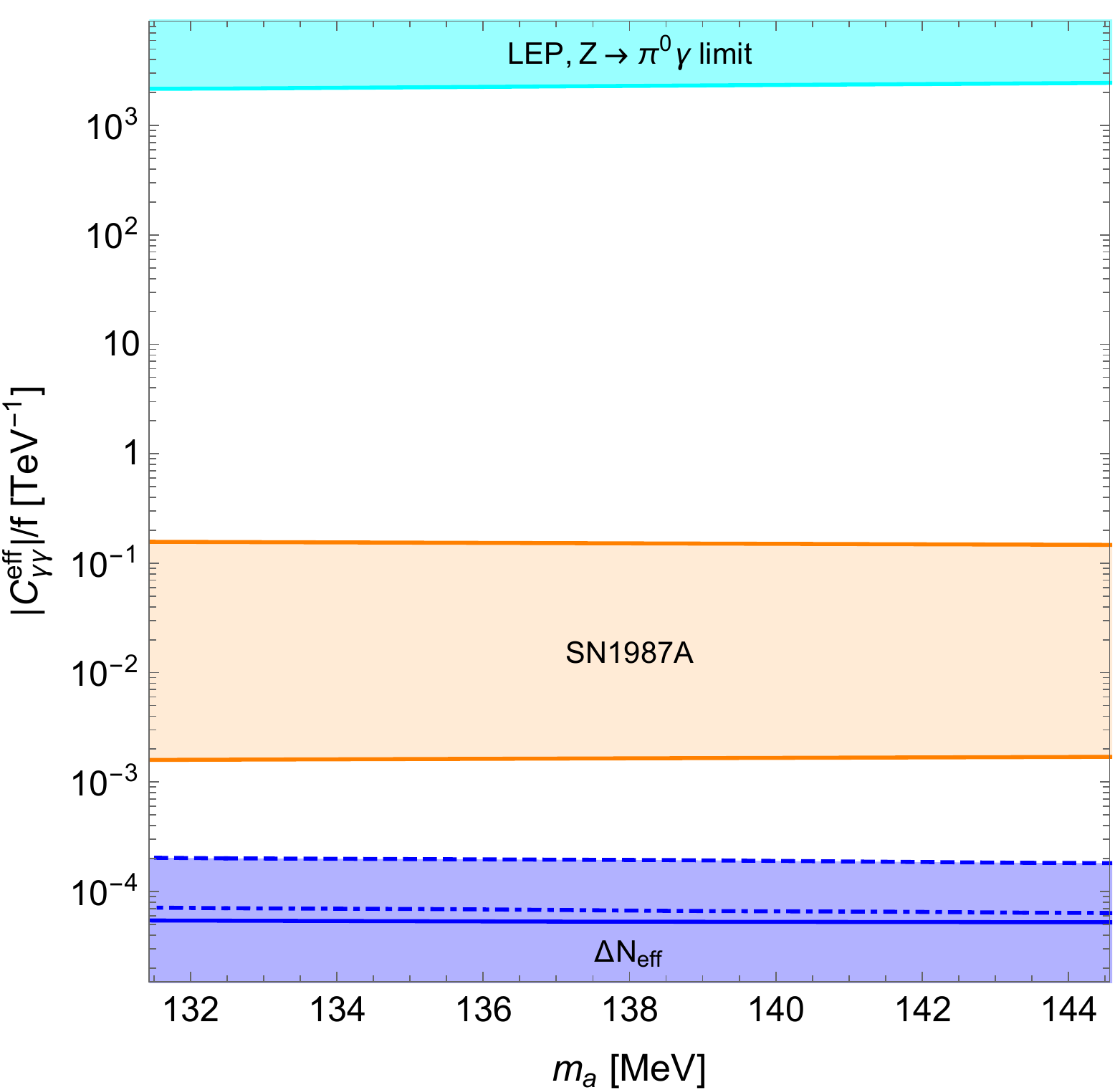}
\caption{The {\it sweetest} spots in the ALP parameter space, having the highest sensitivity to prove the flavorful ALP below $\sim$ GeV, which are presently loopholes (blank domains). 
Current bounds are set on $C_{\gamma \gamma}^{\rm eff}$ normalized to the ALP decay constant $f$, defined in Eq.~(\ref{a-gamma-gamma}), for the ALP mass $m_a \sim 140$ MeV. 
The cyan shaded area is excluded by a conservative upper limit on $Z \to \pi^0 + \gamma$, which has currently been most stringently placed by the LEP searches~\cite{Acciarri:1995gy,Aaltonen:2013mfa,Jaeckel:2015jla,Bauer:2017ris}, without taking the $\pi^0$ contamination with the 140 MeV ALP into account. 
The orange and blue shaded domains are ruled out by SN1987A \cite{Masso:1995tw} and the observation of the effective degrees of freedom for relativistic particles today, $\Delta N_{\rm eff}$~\cite{Depta:2020wmr}, respectively. 
Here, the bound from $\Delta N_{\rm eff}$ has been set at 95\% C.L.. The limits from SN1987 and $\Delta N_{\rm eff}$ assume the ALP dominantly coupled to diphoton. 
Bounds from the beam dump experiments are currently inapplicable in the displayed mass range due to the serious contamination with the neutral pion backgrounds.}
\label{fig:summary}
\end{figure}

Observations of supernova~\cite{Jaeckel:2017tud} are free from the pion contamination, and provide bounds on the ALP-photon coupling (see, e.g.,~\cite{Cadamuro:2011fd,Millea:2015qra}). 
Very recently, it has also been pointed out~\cite{Depta:2020wmr} that the effective degrees of freedom for relativistic particles today and possible effects on the Big Bang Nucleosynthesis can give a significant limit for ALPs, and the coupling to photon at around 140 MeV is severely bounded from below. 
However, still, the loopholes will keep open (see Fig.~\ref{fig:summary}). 

One should realize, on the other side of the same coin, that the existence of the loopholes may provide us with ``sweet spots" in the ALP search: 
As seen in Fig.~\ref{fig:summary} such an ALP is to be short-lived enough to decay to diphoton inside detectors, as is the case of the neutral pion. In this sense, $B^{0,\pm} \to K^{0,\pm} +a$ search cannot constrain the ALP. 
Thanks to this fact, the ALP-fermion coupling can be enlarged to enhance other observables. 
Among meson decays, in particular, $B^{0,\pm} \to K^{0,\pm} +a$ is the severest~\cite{Bjorkeroth:2018dzu}, hence the ALP contributions to presently unexplored $b \to s$ transition amplitudes are maximized at around 140 MeV for the overall mass range $\lesssim $ GeV. 
Note also that when the mass goes off the target scale (i.e. above $B$ meson mass scale), the sensitivity to the $(g-2)_\mu$ gets lower and lower, due to the decoupling effect~\cite{Bauer:2019gfk}. 
Therefore, the ``sweet spot" can be indeed {\it sweetest} for a generic ALP to have the highest discovery sensitivity, and would give us the biggest chance to hunt the ALP responsible for the $(g-2)_\mu$ as well. 
Also, it would be expected to provide us with a nontrivial correlation between quark and lepton flavor physics in the presence of the ALP. 

In this paper, we propose the combined measurements of the time-dependent CP asymmetry in $B^0 \to K_S^0 \pi^0 \gamma$ and $B_s \to e^\pm \mu^\mp$ decay, along with the $(g-2)_\mu$, as the new experimental probes having the highest sensitivity for the flavorful ALPs at around 140 MeV. 
Remarkably enough, the deviation of the time-dependent CP asymmetry from the 
SM prediction can be seen, whichever the $(g-2)_\mu$ is consistent with the SM, or not. 
The proposed complementary measurements can precisely determine the 140 MeV ALP coupling to photon, will fully cover the present loopholes as well, and hence, make a significant help to explore all the parameter space of the ALP. 

\section{ALP couplings}

We begin by introducing a generic ALP model. 
The ALP ($a$) couplings to quarks and leptons, involving the decay constant $f$, arise generically as vector or axialvector current forms, because of the inherit Nambu-Goldstone boson nature. 
They can be written without loss of generality as 
\begin{align}
\mathcal{L}_{a ff} = \frac{\partial_{\mu} a}{2 f} &\left[ (g^d_V)_{ij} \bar{d}_i \gamma^{\mu} d_j + (g^d_A)_{ij} \bar{d}_i \gamma^{\mu} \gamma_5 d_j \right] \nonumber \\
+ \frac{\partial_{\mu} a}{2 f} &\left[ (g^{\ell}_V)_{ij} \bar{\ell}_i \gamma^{\mu} \ell_j + (g^{\ell}_A)_{ij} \bar{\ell}_i \gamma^{\mu} \gamma_5 \ell_j \right]\,,
\label{eq:Lagmass}
\end{align}
where $d_i$ and $\ell_i$ are down type quark and charged lepton fields, with $i$ being the generation index ($i=1,2,3$), the couplings $g^f_{V,A}$ are hermitian matrices, and we have disregarded couplings to up type quark and neutrinos, which are irrelevant to our current proposal. 
The ALP $a$ generically couples also to the photon like 
\begin{align}
\mathcal{L}_{a \gamma \gamma} = C_{\gamma \gamma}^{\rm eff} \frac{\alpha}{4 \pi} \frac{a}{f} F_{\mu \nu} \tilde{F}^{\mu \nu}\,, 
\label{a-gamma-gamma}
\end{align}
with $\alpha$ being the fine structure constant of the electromagnetic coupling, and 
the inclusive effective coupling denoted as $C_{\gamma \gamma}^{\rm eff}$ including both direct and indirect $a \mathchar`-\gamma \mathchar`-\gamma$ vertices, the latter of which can be induced from charged fermion loops. 

We focus on the following ALP couplings, $(g^d_{V, A})_{23}$, $(g^d_A)_{33}$, $(g^{\ell}_{V, A})_{12}$, $(g^{\ell}_A)_{22}$, and $C_{\gamma \gamma}^{\rm eff}$, which are relevant to the existing constraints from flavor observables at the ALP mass around 140 MeV, such as $B_s$-$\overline{B_s}$ mixing ($C_{B_s}$ defined in Eq.~\eqref{CBs}), the CP asymmetry in $B^0 \to K_S^0 \pi^0 \gamma$ ($S_{CP}$ defined in Eq.~\eqref{eq:SCP}), $B_s \to e^\pm \mu^\mp$, $\Upsilon$ decays to $\gamma a$ and $\mu^+ \mu^-$, leptonic $B_s$ meson decays to $\mu^+ \mu^-$ and $e^\pm \mu^\mp$, $(g-2)_{\mu}$ ($\Delta a_{\mu} = a_{\mu}^{\rm exp} - a_{\mu}^{\rm SM}$), and $\mu \to e \gamma$.

\section{Time-dependent CP asymmetry in $B^0 \to K_S^0 \pi^0 \gamma$ } 

Among them, as it turns out later, the CP asymmetry in $B^0 \to K_S^0 \pi^0 \gamma$ $(S_{CP})$ is the most important observable, because it can be deviated from the SM prediction almost irrespective to others. 
Thereby, in this section we discuss the physics on $S_{CP}$ by particularly paying our attention to the ALP contribution forms arising from the Lagrangian (Eqs.~\eqref{eq:Lagmass} and \eqref{a-gamma-gamma}) at loop level. 
For the other observables, the details are shown in Appendix~\ref{AppendixA}. 

The time-dependent CP asymmetry in $B^0 \to K_S^0 \pi^0 \gamma$ decay $S_{CP}$ is evaluated as~\cite{Haba:2015gwa,Kou:2018nap}
\begin{align}
S_{CP} &\equiv \frac{{\rm Im} \left[ e^{- 2 i \beta_{\rm CKM}} \left( C_7^\ast C'_7 + C_7 C'_7{}^{\ast} \right) \right]}{|C_7|^2 + |C'_7|^2}\,, \label{eq:SCP}
\end{align}
where the Wilson coefficients $C^{(')\rm NP}_7$ for the magnetic operators, $O^{(')}_7 = (e / 16 \pi^2) m_b \left( \bar{s} \sigma^{\mu \nu} P_{R(L)} b \right) F_{\mu \nu}$, where $P_{R/L}$ denote the chiral projection operators. 
In Eq.~\eqref{eq:SCP} $C^{(')}_7 = C^{(') {\rm SM}}_7 + C^{(') {\rm NP}}_7$, and $2 \beta_{\rm CKM} \approx 43^{\circ}$ is a CP phase in $B^0 \to K_S^0 \pi^0 \gamma$. 

In the present analysis, we use the SM Wilson coefficients taken from Ref.~\cite{Altmannshofer:2008dz}: $C^{\rm SM}_7 = -0.304$ and $C'^{\rm SM}_7 = -0.006$ at $\mu = 4.8$ GeV. 
The new physics (NP) terms arising from the ALP loops are evaluated, from Eqs.~\eqref{eq:Lagmass} and \eqref{a-gamma-gamma}, as 
\begin{align}
C^{(') {\rm NP}}_7 = C^{(') {\rm NP, \, arch}}_7 + C^{(') {\rm NP, \, BZ}}_7\,. 
\label{C-arch}
\end{align}
Here $C^{(') {\rm NP, \, arch}}_7$ correspond to contributions from the ALP exchange (arch) graphs at the one-loop level, 
\begin{align}
C^{\rm NP, \, arch}_7 = \frac{\sqrt{2}}{4 G_F} \frac{1}{V_{ts}^{\ast} V_{tb}} \frac{1}{6} &\left[ g_s^{k 2} g_s^{k 3} I_{k,1}^{++} + g_p^{k 2} g_p^{k 3} I_{k,1}^{+-} + i \left( g_p^{k 2} g_s^{k 3} I_{k,1}^{-+} - g_s^{k 2} g_p^{k 3} I_{k,1}^{--} \right) \right]\,, \\
C'^{\rm NP, \, arch}_7 = \frac{\sqrt{2}}{4 G_F} \frac{1}{V_{ts}^{\ast} V_{tb}} \frac{1}{6} &\left[ g_s^{k 2} g_s^{k 3} I_{k,1}^{++} + g_p^{k 2} g_p^{k 3} I_{k,1}^{+-} - i \left( g_p^{k 2} g_s^{k 3} I_{k,1}^{-+} - g_s^{k 2} g_p^{k 3} I_{k,1}^{--} \right) \right]\,, 
\end{align}
where $V_{tb}$ and $V_{ts}$ are the Cabibbo-Kobaashi-Maskawa (CKM) matrix elements and $G_F$ is the Fermi constant, and 
\begin{align} 
g_s^{ij} &\equiv - \frac{i }{2} (m_{d_i} - m_{d_j}) \frac{(g^d_V)_{ij}}{f} \, , \notag \\ 
g_p^{ij} &\equiv \frac{1 }{2} (m_{d_i} + m_{d_j}) \frac{(g^d_A)_{ij}}{f} . 
\end{align}
The loop integrals $I_{k,1}^{\pm\pm}$ can be found in the Appendix of Ref.~\cite{Lindner:2016bgg}, and $k$ denotes the quark flavor flowing in the loop, where $k = 1, 2$ and $3$ correspond to the contribution of down, strange and bottom quarks, respectively. 
Since $g_{s(p)}^{ij}$ is proportional to the difference between (sum of) the masses of $i$- and $j$-th generations of down type quark, the dominant contribution comes from the bottom quark loop, i.e. $k = 3$. 
Noting also $g_s^{ii}=0$, we thus find that the dominant part of $C^{(') {\rm NP, \, arch}}_7$ depends on three couplings: $(g^d_V)_{32}$ ($= (g^d_V)_{23}^{\ast}$), $(g^d_A)_{32}$ ($= (g^d_A)_{23}^{\ast}$) and $(g^d_A)_{33}$ ($= (g^d_A)_{33}^{\ast}$). 
It is then clear to see that $C^{(') {\rm NP, \, arch}}_7 = 0$ when $(g^d_A)_{33} = 0$. 

The coupling $C_{\gamma \gamma}^{\rm eff}$ in Eq.~\eqref{a-gamma-gamma} also contributes to $b \to s \gamma$ processes, through the Barr-Zee (BZ) type loops~\cite{Barr:1990vd}. 
This contribution, denoted as $C^{(') {\rm NP, \, BZ}}_7$ in Eq.(\ref{C-arch}), can be evaluated in a way analogous to the BZ type correction to the $\mu \to e \gamma$ amplitude in Refs.~\cite{Marciano:2016yhf,Cornella:2019uxs}. 
This is given as follows: 
\begin{align}
C^{\rm NP, \, BZ}_7 &= \frac{\sqrt{2}}{4 G_F} \frac{1}{V_{ts}^{\ast} V_{tb}} \frac{1}{6} \frac{\alpha}{4 \pi f^2} C_{\gamma \gamma}^{\rm eff} g_{\gamma} (x_b) \Bigl[ (g_V^d)_{32} + (g_A^d)_{32} \Bigr] \,, \\
C'^{\rm NP, \, BZ}_7 &= \frac{\sqrt{2}}{4 G_F} \frac{1}{V_{ts}^{\ast} V_{tb}} \frac{1}{6} \frac{\alpha}{4 \pi f^2} C_{\gamma \gamma}^{\rm eff} g_{\gamma} (x_b) \Bigl[ - (g_V^d)_{32} + (g_A^d)_{32} \Bigr] \,, 
\end{align}
where $x_b \equiv m_a^2 / m_b^2$ and $g_{\gamma} (x)$ is the loop function which is found in the Appendix of \cite{Cornella:2019uxs}.

\section{ALP coupling correlations in flavor observables}
\label{AppendixB}

As to other flavor observables that we are presently interested in, the explicit formulae and the existing limits are presented in Appendix~\ref{AppendixA}. 
To summarize, all the relevant flavor observables are given as functions of the ALP couplings (up to the decay constant factor $1/f$):
\begin{align}
C_{B_s}, \phi_{B_s} &: (g^d_V)_{23}, (g^d_A)_{23} \notag \\
{\rm BR}(\Upsilon \to \gamma a) &: (g^d_A)_{33} \notag \\
{\rm BR}(B_s \to \mu^- \mu^+) &: (g^d_A)_{23}, (g^{\ell}_A)_{22} \notag \\
{\rm BR}(B_s \to e^{\pm} \mu^{\mp}) &: | (g^d_A)_{23} |, c_{e \mu} \notag \\
(g-2)_{\mu} &: (g^{\ell}_A)_{22}, c_{e \mu}, C_{\gamma \gamma}^{\rm eff} \notag \\
{\rm BR}(\mu \to e \gamma) &: (g^{\ell}_A)_{22}, c_{e \mu}, C_{\gamma \gamma}^{\rm eff} \notag \\ 
S_{CP} &: (g^{d}_V)_{23}, (g^{d}_A)_{23}, (g^{d}_A)_{33}, C_{\gamma \gamma}^{\rm eff} . 
\label{coupling-dep}
\end{align} 
We first note that $(g^d_A)_{33}$ can be determined solely by BR$(\Upsilon \to \gamma a)$. 
We also notice that once $(g^d_A)_{23}$ is fixed, we can then determine $(g^d_V)_{23}$ as a function of $C_{B_s}$ and $\phi_{B_s}$. 
The other useful features that we note are BR$(B_s \to e^{\pm} \mu^{\mp}) \propto | (g^d_A)_{23} |^2 c_{e \mu}^2$ and BR$(\mu \to e \gamma) \propto c_{e \mu}^2$. 
Therefore, $| (g^d_A)_{23} |$ and $c_{e \mu}$ can be written as functions of $(g^{\ell}_A)_{22}$, $C_{\gamma \gamma}^{\rm eff}$ and branching ratios of $B_s \to e^{\pm} \mu^{\mp}$ and/or $\mu \to e \gamma$.

\section{Covering the 140 MeV loopholes} 

Although the ALP dominantly decays into $\gamma\gamma$, the diphoton signal cannot be distinguished from the neutral pion signal, so that there are loopholes at around the pion mass of 140 MeV~\cite{Kitahara:2019lws} (and also see Fig.~\ref{fig:summary}). 
However, such loopholes can be resolved and covered by measurements on the time-dependent CP asymmetry in $B^0 \to K_S^0 \pi^0 \gamma$ ($S_{CP}$), and $B_s \to e^\pm \mu^\mp$, together with the $(g-2)_\mu$ ($\Delta a_\mu$), that is our main proposal, and will be demonstrated below shortly. 

As one reference point, we may fix the flavor observable values as follows: 
\begin{align}
C_{B_s} &= 1.11\,, \notag \\
\frac{{\rm BR}(B_s \to \mu^- \mu^+)_{\rm exp}}{{\rm BR}(B_s \to \mu^- \mu^+)_{\rm SM}} &\simeq 0.822\,, \notag \\
\Delta a_{\mu} &= 2.61 \times 10^{-9}\,, \notag \\
{\rm BR}(\mu \to e \gamma) &= 6 \times 10^{-14}\,, \notag \\
\frac{{\rm BR}(\Upsilon \to \gamma a)}{{\rm BR}(\Upsilon \to \mu \mu)} &\simeq 1.81 \times 10^{-4}\,,
\label{eq:bench}
\end{align}
where the measured values have been set to those central values, and upper limits have been taken from the future prospects as benchmark values, which are given in Appendix~\ref{AppendixA}. 
Note that according to Ref.~\cite{Endo:2020mev}, the upper bound on $c_{e \mu}$ from the muonium-antimuonium oscillation can be read as
\begin{align}
c_{e \mu} \lesssim 0.45 \left( \frac{m_a}{140 \, {\rm MeV}} \right).
\label{eq:mmosc}
\end{align}
This $c_{e \mu}$ is mostly sensitive to ${\rm BR}(B_s \to e^{\pm} \mu^{\mp})$, because it is solely scaled by $|c_{e \mu}|^2$ (see Eq.(\ref{coupling-dep})). 
In order to be consistent with this bound {\color{black} and Eq.~(\ref{eq:bench})}, we therefore set 
\begin{align} 
{\rm BR}(B_s \to e^{\pm} \mu^{\mp}) = 10^{-12}.
\label{Bs-mue:upper}
\end{align}
This is actually taken to be the upper bound for the present analysis: 
We have checked that when BR$(B_s \to e^{\pm} \mu^{\mp}) \gtrsim 10^{-12}$, there exists no solution which satisfies Eq.~\eqref{eq:mmosc}. 
Thus, we can evaluate the time-dependent CP asymmetry parameter $S_{CP}$ as a function of $C_{\gamma \gamma}^{\rm eff}$. 
We see the following features: 

\begin{itemize} 

\item As seen from Eq.~\eqref{eq:SCP} the structure of new physics term in $S_{CP}$ is shared by two contributions proportional to the CP phase in the SM, or new physics phase, which currently come from the presence of the imaginary part of ALP coupling $(g^d_R)_{23}$. 
Actually, a sizable deviation from the SM prediction ($C'^{\rm NP}_7 =\mathcal{O}(0.001)$) can be generated even when $C'^{\rm NP}_7$ has no imaginary part (i.e. solely with the KM phase in the SM). 
The deviation of our prediction has been enhanced by nonzero imaginary part of $C'^{\rm NP}_7$ coming from Im[$(g^d_R)_{23}$]. 

\item $S_{CP}$ can be large enough when $| C_{\gamma \gamma}^{\rm eff} | > \mathcal{O}(100)$/TeV. 
The stringent constraint from the muonium-antimuonium oscillation in Eq.~\eqref{eq:mmosc} (or Eq.~\eqref{Bs-mue:upper}) gives the significant impact on $S_{CP}$ as a function of $C_{\gamma \gamma}^{\rm eff} / f$. In fact, when the current size of deviation on $\Delta a_\mu$ (the benchmark value in Eq.~\eqref{eq:bench}) is fixed, we find that they are almost stuck to 
\begin{align} 
& S_{CP} \simeq -0.0230, \quad {\rm or} \quad -0.0316 \notag \\ 
{\rm at} \qquad & | C_{\gamma \gamma}^{\rm eff} / f | \simeq 209/ {\rm TeV} \,,
\label{eq:SCPg-2neq0}
\end{align} 
which are compared with the SM prediction, $S_{CP}^{\rm SM} \simeq -0.0269$~\cite{Ball:2006eu}. 

\item We also found that when $\Delta a_{\mu}$ deviates from the benchmark value in Eq.~\eqref{eq:bench}, the predicted value of $C_{\gamma \gamma}^{\rm eff}$ is changed to get larger monotonically with increasing $\Delta a_{\mu}$ and vice verse. 
In this sense, it is possible to explore the loopholes in Fig.~\ref{fig:summary} in light of the fate of $\Delta a_{\mu}$ in the future. 
Particularly, persistence of nonzero $\Delta a_{\mu}$ will make $S_{CP}$ cover the blank range determined by the upper part of the SN1987A constraint and the lower part of the LEP limit. 

\end{itemize}


\begin{figure}[ht]
\begin{center}
\includegraphics[width=0.45\textwidth,bb= 0 30 450 288]{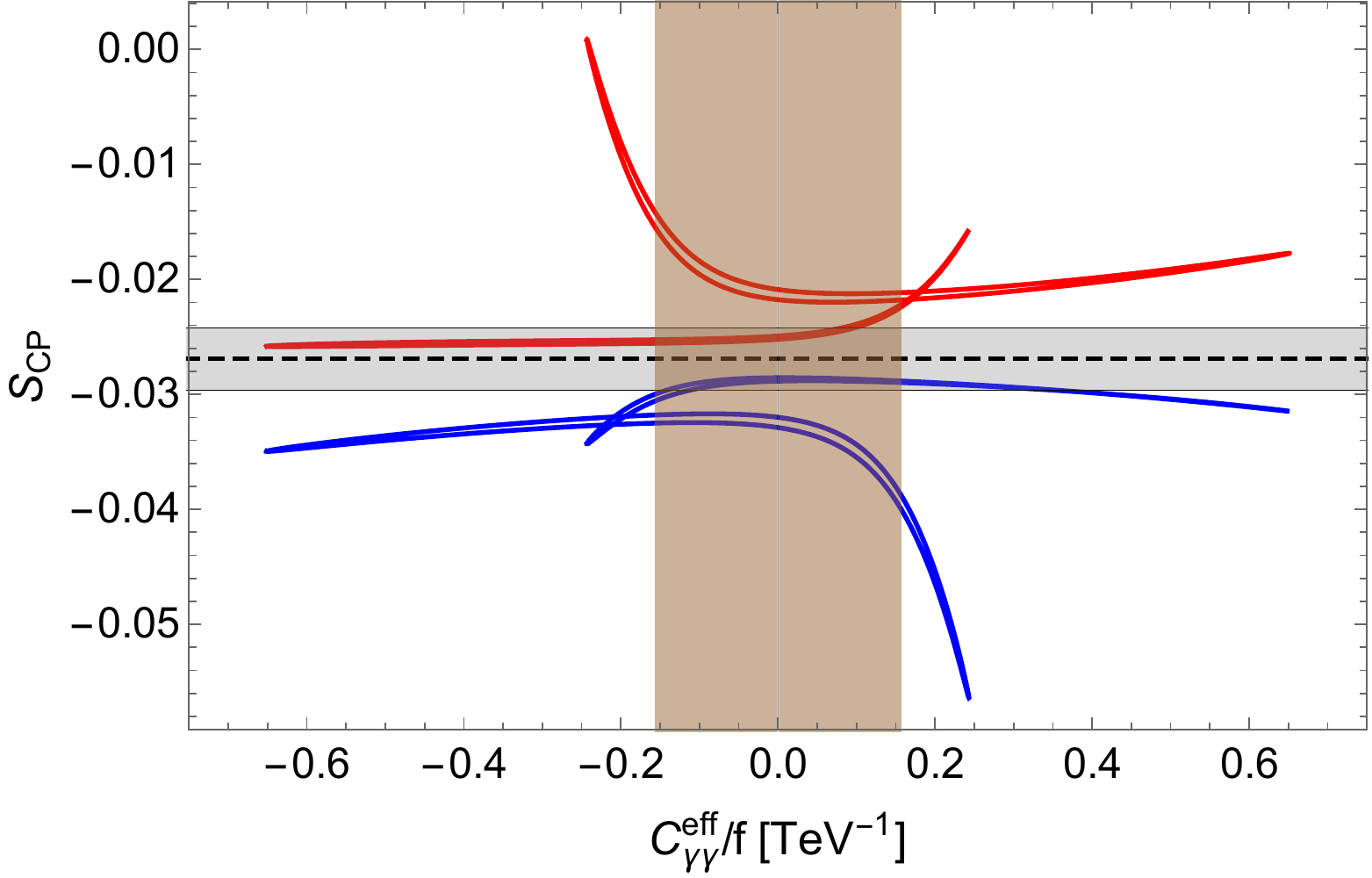}
\end{center}
\caption{$S_{CP}$ versus $C_{\gamma \gamma}^{\rm eff}/f$ for $\Delta a_\mu=0$, with 
other flavor observables fixed as shown in the main text, satisfying the current experimental limits including the one solely for the $C_{\gamma\gamma}^{\rm eff}$ in Eq.~\eqref{eq:Cggconst}. 
The red and blue lines have been created for the same size of couplings, but the different overall sign of $(g^d_V)_{23}$. 
Horizontal black dashed line shows the SM prediction ($S_{CP}^{\rm SM} \simeq -0.0269$~\cite{Ball:2006eu}). 
The gray shaded regions show the deviation from the SM prediction by taking it to be $< 10$\%. 
The domain surrounded by the brown vertical lines are excluded (see Eq.~\eqref{eq:Cggconst}).}
\label{fig:SCPpftr}
\end{figure}

Another interesting reference points to the case where $\Delta a_{\mu} = 0$ in our future. 
The result for $S_{CP}$ is shown in Fig.~\ref{fig:SCPpftr}. 
In this case, BR$(B_s \to e^{\pm} \mu^{\mp})$ can be larger than $10^{-12}$ without conflicting with Eq.~\eqref{eq:mmosc}, and hence, we set 
\begin{align}
{\rm BR}(B_s \to e^{\pm} \mu^{\mp}) = 8 \times 10^{-10}\,,
\end{align}
as a benchmark value, which corresponds to the prospected upper bound reported from the LHCb~\cite{Bediaga:2018lhg}. 
The other input values are set to the same ones as in Eq.~\eqref{eq:bench}, except for $\Delta a_{\mu}$. 
Then the following features are detected and deduced: 

\begin{itemize} 

\item The deviation becomes larger than the case with the current central value of $\Delta a_{\mu}$. 
This is because $| (g^d_{V,A})_{23} |$ is allowed to be larger in this case: 
First of all, a small $| C_{\gamma \gamma}^{\rm eff} |$ is favored to realize the benchmark values in Eq.~(\ref{eq:bench}) with $\Delta a_{\mu} = 0$. 
Hence $(g^{\ell}_A)_{22}$ is also desired to be smaller for $\Delta a_{\mu} = 0$. 
This implies that dominant contributions to the $S_{CP}$ from the ALP exchange loops (along with $| (g^d_{V,A})_{23} |$) and the BZ type loops (with $C_{\gamma \gamma}^{\rm eff}$ coupling) are interchanged. 
A large $(g^d_A)_{23}$ is required to achieve the benchmark value of $B_s \to \mu^- \mu^+$ in Eq.~\eqref{eq:bench}. 
For such a large $(g^d_A)_{23}$, the values of $C_{B_s}$ and $\phi_{B_s}$ are realized by a cancellation between $(g^d_V)_{23}$ and $(g^d_A)_{23}$, which results in a large $(g^d_V)_{23}$ as well. 
Thus both $| (g^d_{V,A})_{23} |$ with $\Delta a_{\mu} = 0$ become larger by about two orders of magnitude than those with $\Delta a_{\mu} = 2.61 \times 10^{-9}$. 
Therefore a larger $S_{CP}$ can potentially be predicted with a smaller $C_{\gamma\gamma}^{\rm eff}$ which, with opposite sign, destructively contributes against the dominant ALP exchange loops along with $| (g^d_{V,A})_{23} |$. 

\item For such a small $C_{\gamma \gamma}^{\rm eff}$ $c_{e \mu}$ has to also be small to destructively interfere and realize $\Delta a_{\mu} = 0$. 
Indeed we have $c_{e \mu} = \mathcal{O}(10^{-2})$, which satisfies the muonium-antimuonium oscillation bound in Eq.~\eqref{eq:mmosc}. 

\end{itemize} 

\begin{figure}[t]
\begin{center}
\includegraphics[width=0.45\textwidth,bb= 0 30 450 302]{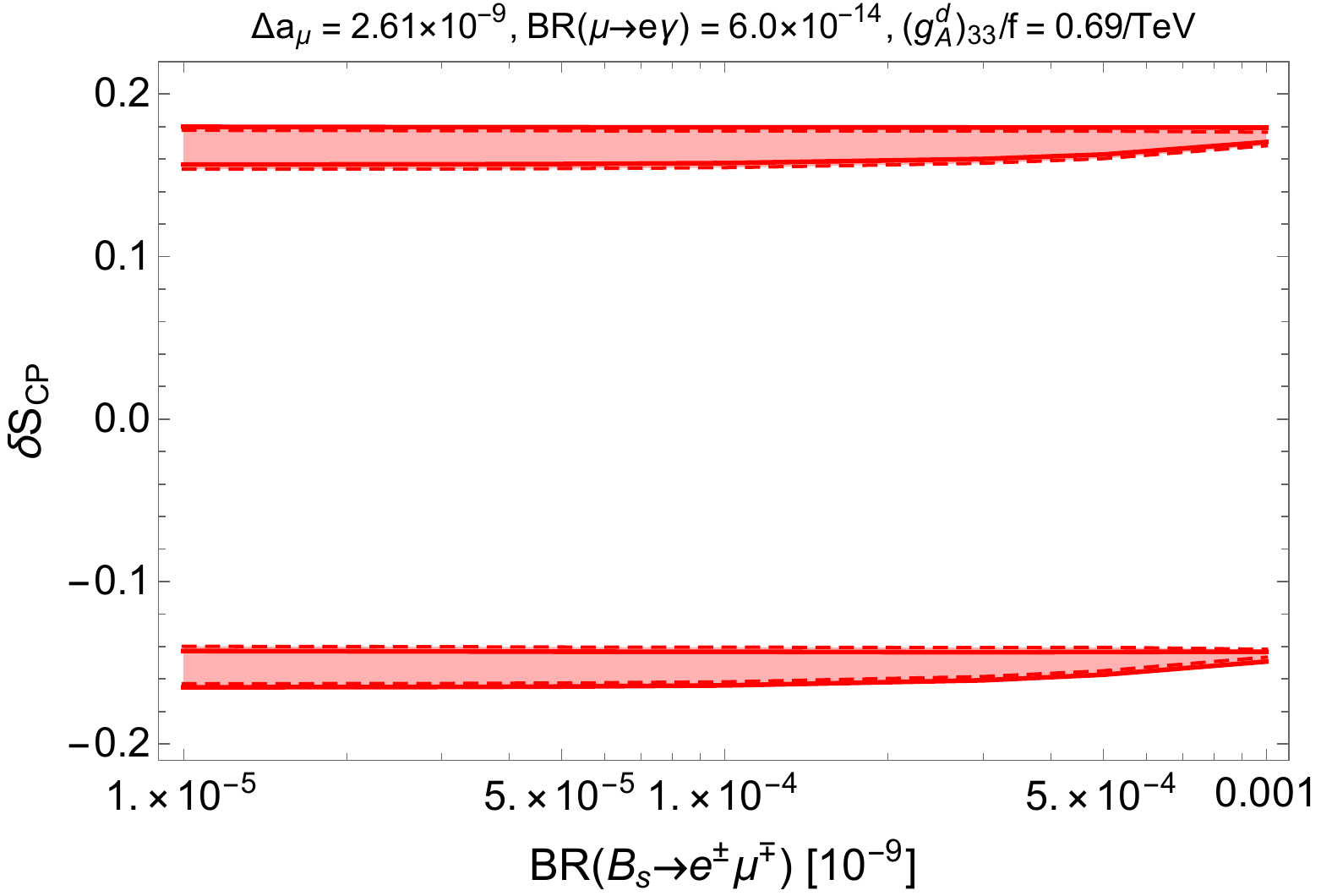}
\end{center}
\caption{The 140 MeV ALP predictions to deviations in $S_{CP}$ ($\delta S_{CP} \equiv S_{CP}^{\rm pred} / S_{CP}^{\rm SM} - 1$) and $B_s \to e^\pm \mu^\mp$ from the SM prediction, for the estimated $\Delta a_{\mu} = 2.61 \times 10^{-9}$ (red shaded domains). 
The solid (dashed) red lines correspond to $C_{\gamma \gamma}^{\rm eff} \geq 0 \, (< 0)$. 
The red shaded bands have been created by choosing $C_{\gamma \gamma}^{\rm eff}$ in an allowed region, in Eq.~(\ref{eq:Cggconst}).}
\label{fig:Bstrendc}
\end{figure}
\begin{figure}
\begin{center}
\includegraphics[width=0.45\textwidth,bb= 0 30 450 314]{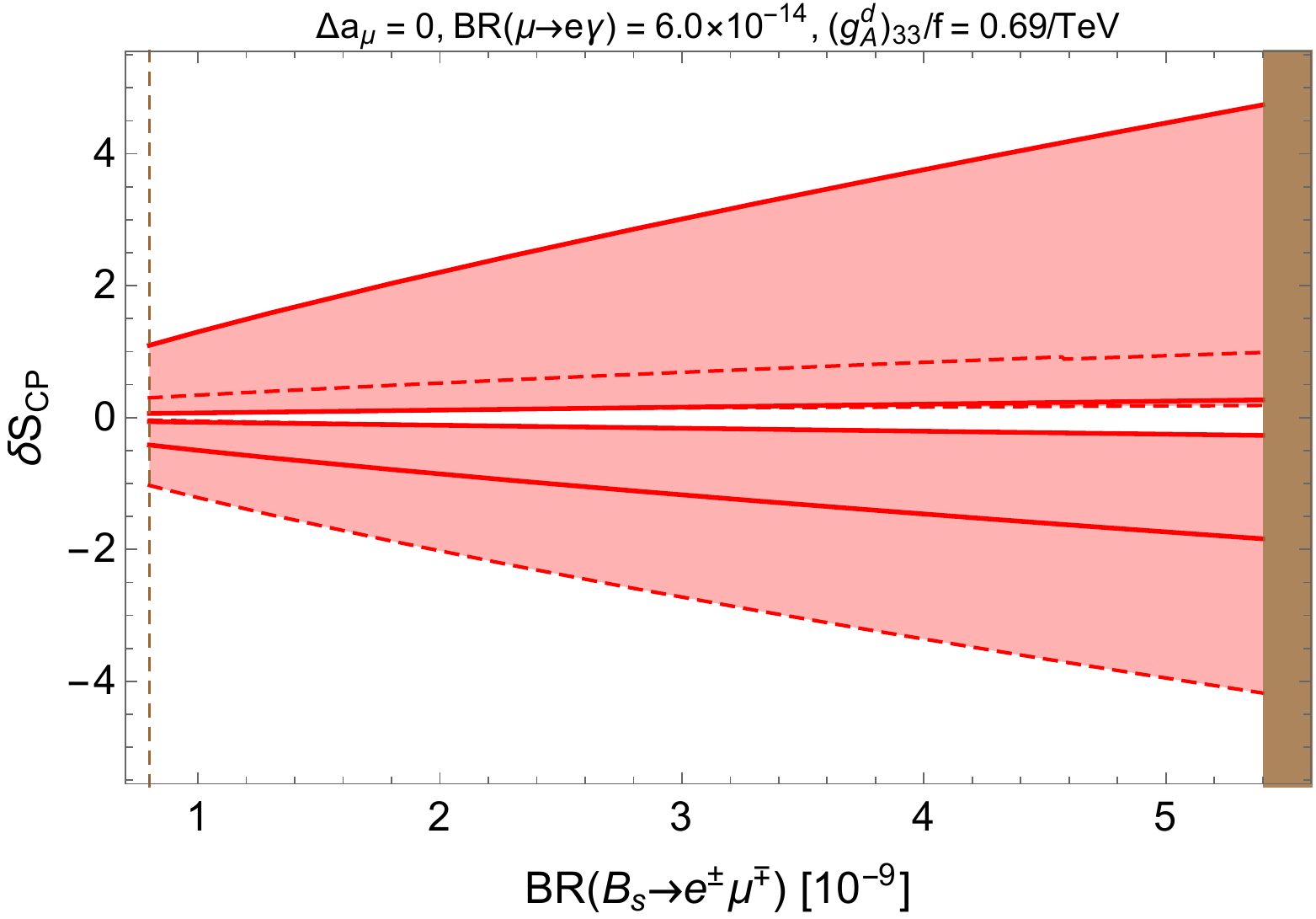}
\end{center}
\caption{The same as Fig.~\ref{fig:Bstrendc} for $\Delta a_{\mu} = 0$. 
The brown shaded region on the right side is excluded at 90\% C.L. \cite{Aaij:2017cza}, and the brown dashed line shows the future prospect for the $B_s \to e^\pm \mu^\mp$ at LHCb \cite{Bediaga:2018lhg}.}
\label{fig:Bstrendf}
\end{figure}

Varying the reference value for BR($B_s \to e^\pm \mu^\mp$), we may also evaluate the correlation with $S_{CP}$. 
See Fig.~\ref{fig:Bstrendc} for $\Delta a_{\mu} = 2.61 \times 10^{-9}$ and Fig.~\ref{fig:Bstrendf} for $\Delta a_{\mu} = 0$. 
In these figures we have kept using the benchmark values except for BR($B_s \to e^\pm \mu^\mp$). 
The red shaded bands have been created by choosing $C_{\gamma \gamma}^{\rm eff}$ in an allowed region, in Eq.~(\ref{eq:Cggconst}).
For $\Delta a_{\mu} \neq 0$ case, $\delta S_{CP}$ is constrained in two narrow bands around $\delta S_{CP} \sim 0.17$ and $\delta S_{CP} \sim -0.15$, and these are almost independent of the value of BR($B_s \to e^\pm \mu^\mp$). 
This can be understood as follow: 
Because of the tiny value of BR($B_s \to e^\pm \mu^\mp$), $(g_A^d)_{23}$ is smaller than $(g_V^d)_{23}$. 
In this case, both $C_7^{(')NP}$ depend highly on the size of $(g_V^d)_{23}$. 
Its size is almost determined by $C_{B_s}$, hence is independent of BR($B_s \to e^\pm \mu^\mp$) value. 
For $\Delta a_{\mu} = 0$ case, on the other hand, $S_{CP}$ has a strong sensitivity on BR($B_s \to e^\pm \mu^\mp$), because $(g_A^d)_{23}$ gives contributions to $S_{CP}$ on the same order as $(g_V^d)_{23}$ does, as explained above. 

We can change the benchmark value of BR($\mu \to e \gamma$). 
The expected results can be easily deduced by following the discussion above: 

\begin{itemize} 

\item Note the scaling of $| (g^d_A)_{23} | \propto 1 / \sqrt{{\rm BR}(\mu \to e \gamma)}$. 
For $\Delta a_\mu \neq 0$ including the reference point $\Delta a_{\mu} = 2.61 \times 10^{-9}$, the size of $\delta S_{CP}$ is fairly independent of BR($\mu \to e \gamma$), because of the weak sensitivity of $| (g^d_A)_{23} |$ to $\delta S_{CP}$. 
Hence $\delta S_{CP}$ can be larger than ${\cal O}(10)$\%, no matter how much smaller BR($\mu \to e \gamma$) is going to be. 

\item For $\Delta a_{\mu} = 0$, $\delta S_{CP}$ depends strongly on $| (g^d_A)_{23} |$, and hence, $\delta S_{CP}$ and BR($\mu \to e \gamma$) will negatively be correlated, and $\delta S_{CP}$ gets monotonically larger as BR($\mu \to e \gamma$) gets smaller and smaller. 

\end{itemize} 
So, at any rate, $\delta S_{CP}$ is predicted to be large enough. 
Note that even the largest values displayed in Fig.~\ref{fig:Bstrendf}, $\delta S_{CP} \sim \pm 4$, which correspond to $S_{CP} \sim 0.08$ and $S_{CP} \sim - 0.13$, are consistent with the current experimental limits~\cite{Amhis:2019ckw,Zyla:2020zbs} within $2\sigma$. 

One may also notice that the results in Figs.~\ref{fig:SCPpftr} and \ref{fig:Bstrendf} depend on the size of $(g^d_A)_{33}$, namely, which has been obtained by taking the upper limit on $\Upsilon \to \gamma a$. 
It turns out, however, that even when we choose $(g^d_A)_{33} = 0$, the size of $\delta S_{CP}$ can be larger than $\mathcal{O}(10)$\%. 
As to the case for $\Delta a_\mu = 0$, we have checked that the results in Fig.~\ref{fig:Bstrendc} are almost independent of $(g^d_A)_{33}$. 

Thus, no matter what future we may or may not have with the NP term in $(g-2)_\mu$, or in BR($\mu \to e \gamma$) and/or BR($\Upsilon \to a \gamma$), $\delta S_{CP}$ can be larger than $\mathcal{O}(10)$\%, which would be testable in future experiments. 
Though other new physics candidates might also give a similar contribution to the $S_{CP}$, the complementary flavor observations would distinguish the 140 MeV ALP from others. 
Moreover, if both $S_{CP}$ and BR$(B_s \to e^\pm \mu^\mp)$ are observed, the size of the ALP coupling to photon can be determined from Eq.~\eqref{eq:SCPg-2neq0} and Fig.~\ref{fig:Bstrendc} for $\Delta a_\mu \neq 0$ and Figs.~\ref{fig:SCPpftr} and \ref{fig:Bstrendf} for $\Delta a_\mu =0$. 

Regarding experimental prospects, even at the Belle II with 50 ab${}^{-1}$ data~\cite{Kou:2018nap}, it will still be hard to observe the signal of $S_{CP}$ since the future uncertainty will just go down from the current one $0.3$ to $0.03$, with assuming the current central value $-0.29$. 
This is the value quoted from Ref.~\cite{Kou:2018nap}, which the current analysis follows. The HFLAV experiments report the central value $-0.15 \pm 0.20$~\cite{Amhis:2019ckw}, and Particle Data Group adapts $- 0.78 \pm 0.59 \pm 0.09$~\cite{Zyla:2020zbs}. 
All of them are consistent each other within the one sigma uncertainty. 
The large uncertainty on the SM prediction essentially comes from the uncontrollable long-distance QCD contributions. 
We, however, hope the future experiments to have a potential to give us some hints in searching for 140 MeV ALPs. 
Actually, an interesting possibility to decrease the uncertainty from long-distance contributions has recently been proposed, by combining the observations of the time-dependent CP asymmetry for exclusive $B$ meson decays into vector or axial-vector meson and photon~\cite{Gratrex:2018gmm}. 
This method will be utilized to measure the $S_{CP}$ at Belle II~\footnote{We are grateful to Akimasa Ishikawa for this information.}, and may help decrease the major theoretical uncertainty in the SM prediction and improve the future experimental sensitivity. 
If future experiments could shrink its uncertainty with a central value significantly deviated from the SM prediction, flavorful 140 MeV ALPs might favor to have the $(g-2)_\mu$ consistent with the SM one.

\section{Conclusion and Discussions} 

In conclusion, the discovery potential of the time-dependent CP asymmetry in $B^0 \to K_S^0 \pi^0 \gamma$ decay has the highest sensitivity to probe flavorful ALPs at around 140 MeV. 
The predicted deviation from the SM can be larger than ${\cal O}(10)$\%, and turns out to be fairly irrespective to the presence of new physics in the anomalous magnetic moment of muon, or $\Upsilon$ decays, or $\mu \to e \gamma$, which will in the future fully cover the current {\it sweetest} spots/loopholes. 
Complementary measurements for the CP asymmetry in $B^0 \to K_S^0 \pi^0 \gamma$ and $B_s \to e^\pm \mu^\mp$, in conjunction with the anomalous magnetic moment of muon, can precisely determine the ALP coupling to photon at around 140 MeV, and will give a significant help to fully cover the ALP parameter space including the regions which cannot be explored by existing and prospected direct search experiments. 

Several comments and discussions are in order: 

\begin{itemize} 

\item As to the future prospect regarding the direct ALP production bound in the beam dump experiments, it has recently been reported~\cite{Dusaev:2020gxi,Banerjee:2020fue} that the ALP detection sensitivity can be increased by an improved analysis on the Primakoff scattering with nucleus, but the signal over backgrounds at around 140 MeV gets weak enough and still suffers from the contamination with neutral pions, in contrast to an optimistic prospect in~\cite{Beacham:2019nyx}. 
This contamination problem also involves the photoproduction of ALPs in the PrimEX and GlueEX experiments~\cite{Aloni:2019ruo}. 

\item We have included the collider experimental limit on the ALP, from LEP in Fig.~\ref{fig:summary}, but actually it still suffers from the contamination with the neutral pion background, in a sense similar to the beam dump experiments. 
In Ref.~\cite{Bauer:2017ris}, Higgs-mediated processes such as $h \to Z + a$ and $h \to aa$ as well as $Z \to a + \gamma$ have been discussed and are shown to have high sensitivity to probe the ALP signal for a wide range of the ALP parameter space including the 140 MeV ALP. 
However, those should still suffer from the contamination with the neutral pion background at the ALP mass around 135 -- 140 MeV, which is not argued in the literature.
(Actually, the authors have clearly mentioned “We are not in a position to provide detailed estimates of detector and reconstruction efficiencies, or to perform solid background estimates.” in the paper (the fifth line from bottom, page 32, for the published version). 
Therefore, the prospected plots in Figs.~17 and 23 in the literature cannot reliably be applied to the loophole/the sweetest spots for the flavorful 140 MeV ALP, hence will not fully cover the ALP parameter space, just like the case of beam dump experiments --- Any of direct 140 MeV ALP productions at collider experiments cannot be free from the pion contamination. Our proposal in a view of flavor physics is totally free from the pion background, and leads to the highest sensitivity as the “sweetest spot” in the ALP full parameter space. 
This is how in Fig.~\ref{fig:summary} the existing LEP (and also Tevatron, LHC) limits on the ALP around 140 MeV have been placed by simply applying the upper limit on the SM prediction $Z \to \pi^0 + \gamma$ event, without taking into account the pion contamination, not by isolated ALP signals. 

\item In Ref.~\cite{Alves:2019xpc} it has been addressed that the enhanced monophoton plus a large missing energy signature can cover one of loopholes around 140 MeV above the supernova constraint in Fig.~\ref{fig:summary}. 
However, the other loophole below the supernova lower limit cannot be probed because their ALP cannot be longer-lived by such a scenario construction, while our present proposal can do it. 
Moreover, such a sterile-ALP coupling scenario is somewhat specialized, and beyond our scope keeping generality. 

\begin{figure}[t]
\centering
\includegraphics[width=0.4\textwidth,bb= 0 0 450 438]{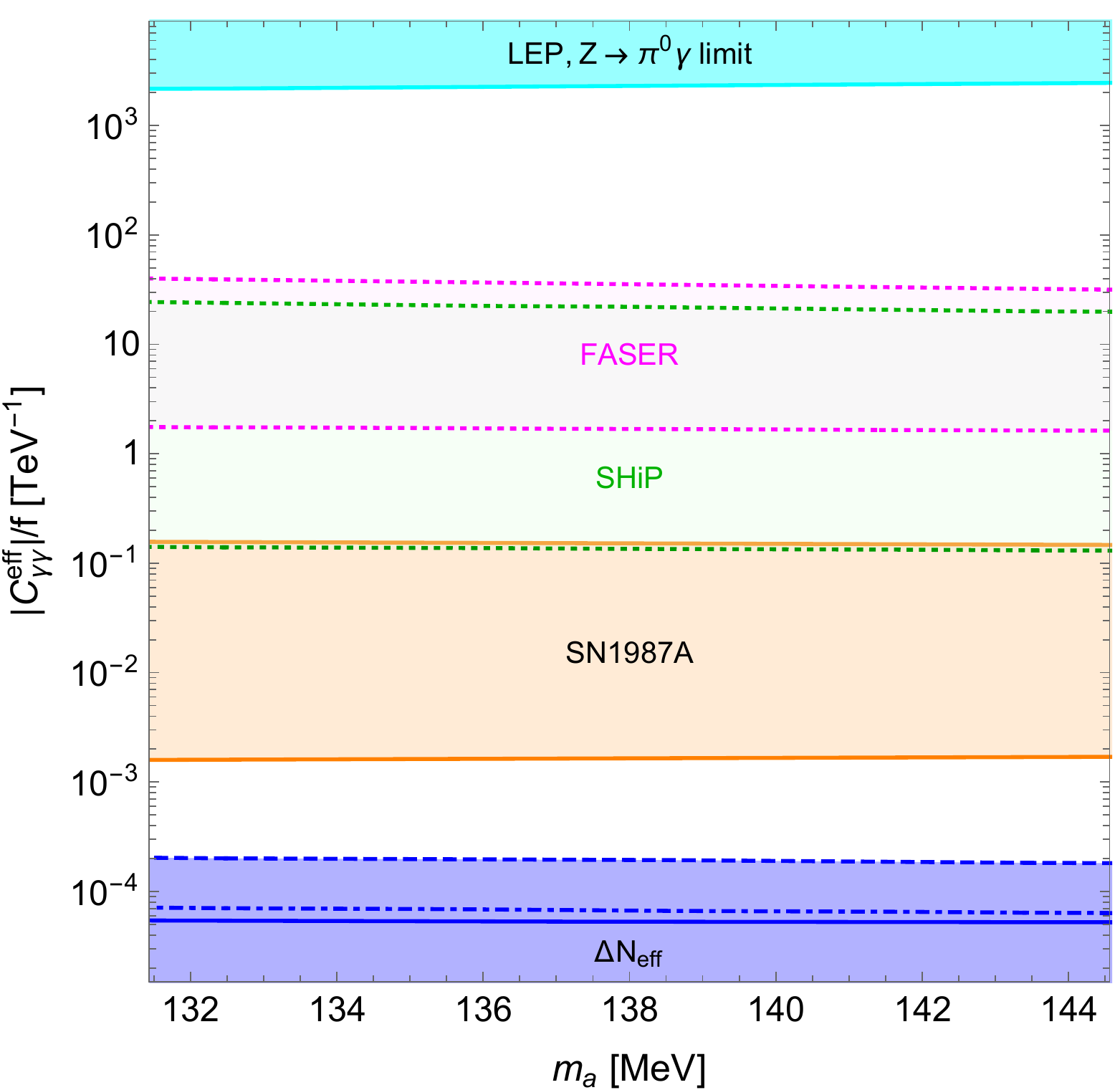}
\caption{Prospected LLP experiment projections onto the ALP parameter space in Fig.~\ref{fig:summary}. 
The characteristic accessibility is the blank domain, below the supernova lower limit and above the FASER prospected upper limit, hence the presently proposed ALP probe will fully cover the 140 MeV ALP parameter space.}
\label{fig1}
\end{figure}

\item 
Comparison with the long-lived particle (LLP) detection experiments (e.g. FASER and SHiP) is shown in Fig.~\ref{fig1}. 
In the figure, we display the prospected sensitivity of the LLP search projected onto the ALP parameter space, by quoting the references (\cite{Feng:2018pew} for FASER and \cite{Alekhin:2015byh} for SHiP). 
Those LLP experiments get less sensitive when the $a-\gamma-\gamma$ coupling is so large that the lifetime (decay length in the experiments) becomes too short to reach the detectors. Actually, as seen from the figure, the ALP cannot be probed when the $a-\gamma-\gamma$ coupling is of ${\cal O}(100/{\rm TeV})$. 
On the other side, the LLP experiments will also fail to detect the ALP when the lifetime gets long enough that the ALP does not decay inside the detector. 
The corresponding boundary will coincide with the upper limit placed by the supernova constraint. 
Thus, the domain surrounded by the lower limit from the supernova and the upper limit from the $\Delta N_{\rm eff}$ will still remain as an unexplored region, which corresponds to the $a-\gamma-\gamma$ coupling of ${\cal O}(10^{-4}/{\rm TeV} - 10^{-3}/{\rm TeV})$. 
Other prospected LLP experiments (e.g. Belle II) have the similar detectability limit (bounded from both above and below). 
Furthermore, to place the bound, those LLP experiments need to assume the ALP coupling to diphoton to be dominant compared to couplings to SM fermions, where the latter would actually be important to discuss ALP significance on flavor physics, as addressed in the manuscript. 
(In particular, the Belle II's paper~\cite{Dolan:2017osp} clearly states that they do not have sensitivity enough to resolute the ALP signal at around the neutral pion mass.) 

\end{itemize}

Our proposal given in the present paper actually covers all the parameter space, by monitoring the ALP contribution to the muon $g-2$, including the two domains that the LLP experiments cannot explore:
the top blank region in the above Fig.~\ref{fig1} is precisely the benchmark regime in which the anomaly on the muon $g-2$ can be explained by the 140 MeV ALP (see Eq.~\eqref{eq:SCPg-2neq0}) while going down to the bottom blank region, one cannot see the significance of ALP in the muon $g-2$ (Fig.~\ref{fig:SCPpftr}). 
Both two cases predict a sizable contribution to the time-dependent CP asymmetry in $B^0 \to K_S^0 \pi^0 \gamma$ and $B_s \to e^\pm \mu^\mp$ decay. 

Thus, our proposal should have the novelty, when compared to the existing and also prospected experiments even including the LLP detection experiments.

With this exclusion or discovery potential at hand, possible implications to the underlying theory for such flavorful axion-like particles (e.g. responsible for the origin of its mass and couplings ) would be worth exploring, as well as the impact on thermal history of universe. 
The region of the favored parameter space as in the figures implies $f \sim \, {\rm TeV} \times (C_{\gamma\gamma}^{\rm eff}/100)$. 
Supposing $C_{\gamma\gamma}^{\rm eff} < {\cal O}(1)$, one gets $f \lesssim 10$ GeV, 
which could thus be much smaller than the typical QCD axion decay constant $(\sim 10^{9-10}\, {\rm GeV})$ or flavon-like axion's~\cite{Ema:2016ops,Calibbi:2016hwq}. 
When one naively embeds the present ALP model into a linear sigma model description such as flavon models, this small $f$ would result in further constraints from four fermion operators among the SM quarks and leptons, induced by the exchange of the radial sigma mode. 
Or, it would turn to a constraint for the presence of some new (hidden) QCD-like dynamics responsible for the 140 MeV ALP, having the intrinsic scale at almost the same scale as QCD, $f \sim 100$ MeV, which implies somewhat smaller $C_{\gamma\gamma}^{\rm eff} \sim 10^{-2}$. 
More conservatively, when the coupling is at most at a perturbative limit, i.e., $C_{\gamma\gamma}^{\rm eff} < {\cal O}(4 \pi)$, we would have $f \sim 100$ GeV. 
At any rate, NP particles with mass on the order of the $f$ constant scale would need to be well secluded from the SM particles, to avoid the existing collider experiment limit. 
This is highly model-dependent, which is beyond the current scope. 
More on this issue is worth pursing elsewhere.

\section*{Acknowledgements} 

The authors would like to thank Akimasa Ishikawa, Teppei Kitahara, 
Satoshi Mishima, Fuminobu Takahashi and Koji Tsumura for helpful discussions and useful comments. 
The authors also thank the Yukawa Institute for Theoretical Physics at Kyoto University. Discussions during the YITP workshop YITP-W-20-08 on "Progress in Particle Physics 2020" were useful to improve this work. 
This work was supported in part by the National Science Foundation of China (NSFC) under Grant No.11747308, 11975108, 12047569 and the Seeds Funding of Jilin University (S.M.). 
H.I. and Y.S. thank for the hospitality of Center for Theoretical Physics and College of Physics, 
Jilin University where the present work has been partially done. 
The work of H.I. was partially supported by JSPS KAKENHI Grant Numbers 18H03708.

\appendix 

\section{ALP Flavor Observables}
\label{AppendixA}

In this Appendix, the flavor observables and constraints relevant to the present ALP study (other than $S_{CP}$) are listed. 
The relevant couplings for the analysis are
\begin{align}
\mathcal{L}_{a ff} &= \frac{\partial_{\mu} a}{2 f} \left[ 
(g^d_V)_{ij} \bar{d}_i \gamma^{\mu} d_j + (g^d_A)_{ij} \bar{d}_i \gamma^{\mu} \gamma_5 d_j \right] + 
\frac{\partial_{\mu} a}{2 f} \left[
(g^{\ell}_V)_{ij} \bar{\ell}_i \gamma^{\mu} \ell_j + (g^{\ell}_A)_{ij} \bar{\ell}_i \gamma^{\mu} \gamma_5 \ell_j \right]\,, \label{eq:Lagmass-app} \\
\mathcal{L}_{a \gamma \gamma} &= 
C_{\gamma \gamma}^{\rm eff} \frac{\alpha}{4 \pi} \frac{a}{f} F_{\mu \nu} \tilde{F}^{\mu \nu}\,. \label{a-gamma-gamma-app}
\end{align}

\subsection{Neutral B meson mixing}

The ALP contribution to the $B_s$-$\overline{B_s}$ mixing including the mass range around 140 MeV has recently been studied~\cite{Bjorkeroth:2018dzu,MartinCamalich:2020dfe}. 
According to Ref.~\cite{MartinCamalich:2020dfe}, the ALP contribution to $\Delta m_{B_s}$ is estimated by using the latest lattice results~\cite{Dowdall:2019bea}. 
We refer readers to the literature for details. 
The resultant form goes like 
\begin{align}
\frac{\Delta m_{B_s}}{m_{B_s}} &= \left| 0.077(8) \,\, {\rm GeV}^2 \left( \frac{(g^d_A)_{23}}{2 f} \right)^2 - 0.020(2) \,\, {\rm GeV}^2 \left( \frac{(g^d_V)_{23}}{2 f} \right)^2 \right|\,.
\label{eq:DelMBs}
\end{align}
The measured value is $\Delta m_{B_s} = 1.1688(14) \times 10^{-8}$ MeV~\cite{Tanabashi:2018oca}, with which the SM prediction is consistent~\cite{Dowdall:2019bea}. 
To find the possible size of new physics (NP) contributions, we use results from the UTFit collaboration~\cite{Bona:2006sa,Bona:2007vi}. 
Then it is convenient to define the following form for the $B_s$ mixing parameters: 
\begin{align}
C_{B_s} e^{2 i \phi_{B_s}} = \frac{\langle B_s^0 | \mathcal{L}_{\rm eff}^{\rm SM + NP} | \overline{B_s}^0 \rangle}{\langle B_s^0 | \mathcal{L}_{\rm eff}^{\rm SM} | \overline{B_s}^0 \rangle}\,.
\label{CBs}
\end{align}
Note that $\Delta m_{B_s} = 2 | \langle B_s^0 | \mathcal{L}_{\rm eff} | \overline{B_s}^0 \rangle |$, and then, the SM prediction points to $C_{B_s} = 1$ and $\phi_{B_s} = 0$. 
By the global fit to CKM observables, we find the best fit values $C_{B_s} = 1.110 \pm 0.090$ and $\phi_{B_s} = (0.60 \pm 0.88)^{\circ}$. 
In the main text, these observables have been used to determine the effective coupling combination $(g^d_V)_{23}/f$ in Eq.~\eqref{eq:Lagmass-app}.

\subsection{Radiative bottomonium decay}

The process $\Upsilon \to \gamma \cancel{\it{E}}_{T}$ was searched by the BaBar, and the current upper limit on the branching ratio is $4.5 \times 10^{-6}$ at 90\% C.L. for the case where the invisible state is a light scalar with mass $m_a < 8$ GeV~\cite{delAmoSanchez:2010ac}. 
This process can be used to constrain the coupling combination $(g^d_A)_{33} / f$ arising from Eq.~\eqref{eq:Lagmass-app}. 
As in Ref.~\cite{Wilczek:1977zn}, the branching ratio normalized to BR($\Upsilon \to \mu \mu$) can be estimated as 
\begin{align}
\frac{{\rm BR}(\Upsilon \to \gamma a)}{{\rm BR}(\Upsilon \to \mu \mu)} = \frac{m_b^2}{2 \pi \alpha} \left( \frac{(g^d_A)_{33}}{f} \right)^2\,. 
\end{align}

When we use the experimental value of BR$(\Upsilon \to \mu \mu) = 2.48 \times 10^{-2}$~\cite{Tanabashi:2018oca} assuming negligible ALP corrections, the upper bound on the $(g^d_A)_{33} / f$ can be read as 
\begin{align}
\left| \frac{(g^d_A)_{33}}{f} \right| < \frac{0.69}{\rm TeV}\,.
\label{eq:constUpdec}
\end{align}

\subsection{Leptonic B meson decays}

The $B_s \to \ell_i \bar{\ell}_j$ decay width can be estimated by the couplings in the Lagrangian Eq.~\eqref{eq:Lagmass-app} as 
\begin{align}
\Gamma (B_s \to \ell_i \bar{\ell}_j) &= \frac{m_{B_s}^3 f_{B_s}^2}{128 \pi} \frac{[\lambda (1, r_i^2, r_j^2)]^{1/2}}{(1 - r_a^2)^2} \left| \frac{(g^d_A)_{23}}{f} \right|^2 \label{eq:decBstoell} \\
&\hspace{3.0em}\times \Biggl\{ \Biggr. \left| \frac{(g^{\ell}_V)_{ij}}{f} \right|^2 (r_i - r_j)^2 \left[ 1 - (r_i + r_j)^2 \right] + \left| \frac{(g^{\ell}_A)_{ij}}{f} \right|^2 (r_i + r_j)^2 \left[ 1 - (r_i - r_j)^2 \right] \Biggl. \Biggr\}\,, \nonumber
\end{align}
where $m_{B_s}$ and $f_{B_s}$ are the mass and decay constant of $B_s$ meson, $r_i \equiv m_{\ell_i} / m_{B_s}$ with $m_{\ell_i}$ being a mass of $i$-th generation of charged lepton, and $r_a \equiv m_a / m_{B_s}$. 
Here, $\lambda (x, y, z) = x^2 + y^2 + z^2 - 2 x y - 2 y z - 2 z x$. 
Note that this decay width is symmetric under $r_i \leftrightarrow r_j$. 

For $B_s \to \mu^- \mu^+$ decay, we should consider the interference between the SM and NP contributions. 
Therefore, we use the generic form of branching ratio given in Ref.~\cite{Altmannshofer:2011gn}:
\begin{align}
\frac{{\rm BR}(B_s \to \mu^- \mu^+)}{{\rm BR}(B_s \to \mu^- \mu^+)_{\rm SM}} = |S|^2 \left( 1 - \frac{4 m_{\mu}^2}{m_{B_s}^2} \right) + |P|^2\,,
\end{align}
where
\begin{align}
S = \frac{m_{B_s}^2}{2 m_{\mu}} \frac{C_S - C'_S}{| C_{10}^{\rm SM} |}, ~~~ P = \frac{m_{B_s}^2}{2 m_{\mu}} \frac{C_P - C'_P}{C_{10}^{\rm SM}} + \frac{C_{10} - C'_{10}}{C_{10}^{\rm SM}}\,.
\end{align}
In the generic ALP model, NP contributions are induced in $C_P^{(')}$, arising as the coefficient of the pseudoscalar current of $m_b ( \bar{s} P_{R(L)} b ) ( \bar{\ell} \gamma_5 \ell )$. 
Thus we focus only on the $C_P - C'_P$ term to get 
\begin{align}
C_P - C'_P = g_{\rm SM}^{-1} \frac{m_{\mu}}{m_{B_s}^2 - m_a^2} \frac{(g^d_A)_{23}}{f} \frac{(g^{\ell}_A)_{22}}{f}\,,
\end{align}
where $g_{\rm SM} = - \frac{4 G_F}{\sqrt{2}} V_{tb} V_{ts}^{\ast} \frac{\alpha}{4 \pi}$, with $V_{tb}$ and $V_{ts}$ being the CKM matrix elements and $G_F$ the Fermi constant. 

The current experimental result and the SM prediction for $B_s \to \mu^- \mu^+$ are~\cite{Tanabashi:2018oca,Bobeth:2013uxa}
\begin{align}
{\rm BR}(B_s \to \mu^- \mu^+)_{\rm exp} &= (3.0 \pm 0.4) \times 10^{-9}\,, \label{eq:expBstomumu} \\
{\rm BR}(B_s \to \mu^- \mu^+)_{\rm SM} &= (3.65 \pm 0.23) \times 10^{-9}\,, \label{eq:SMBstomumu}
\end{align}
form which we note the SM to be consistent with the experimental result within $1.5\sigma$. 

For $B_s \to e^{\pm} \mu^{\mp}$, on the other hand, the branching ratio for SM is negligible due to the absence of lepton flavor violation. 
The current experimental bound is~\cite{Aaij:2017cza}
\begin{align}
{\rm BR}(B_s \to e^{\pm} \mu^{\mp})_{\rm exp} &< 5.4 (6.3) \times 10^{-9} ~~~ (90\% ~ (95\%) ~ {\rm C.L.})\,, \label{eq:expBstoemu}
\end{align}
and the future prospect reported from the LHCb collaboration \cite{Bediaga:2018lhg} is 
\begin{align}
{\rm BR}(B_s \to e^{\pm} \mu^{\mp})_{\rm exp} &< 8 \times 10^{-10}\,.
\label{eq:expBstoemuftr}
\end{align}
Since $r_2 \gg r_1$, the dominant part of the ALP contribution to the branching ratio is evaluated as
\begin{align}
{\rm BR}(B_s \to e^{\pm} \mu^{\mp}) &\simeq \frac{m_{B_s}^3 f_{B_s}^2}{32 \pi \Gamma_{B_s}} \frac{[\lambda (1, r_1^2, r_2^2)]^{1/2}}{(1 - r_a^2)^2} \left| \frac{(g^d_A)_{23}}{f} \right|^2 \frac{c_{e \mu}^2}{f^2} r_2^2\,,
\label{eq:BRBstoemu}
\end{align}
where 
\begin{align} 
c_{e \mu} \equiv \frac{1}{\sqrt{2}} \sqrt{| (g^{\ell}_V)_{12} |^2 + | (g^{\ell}_A)_{12} |^2} .
\label{cemu}
\end{align} 
Note that BR$(B_s \to e^{\pm} \mu^{\mp})$ should include separately both BR$(B_s \to e^- \mu^+)$ and BR$(B_s \to e^+ \mu^-)$, and in the ALP case, BR$(B_s \to e^- \mu^+) =$ BR$(B_s \to e^+ \mu^-)$ since the hermicity gives $| (g^{\ell}_{L, R})_{12} | = | (g^{\ell}_{L, R})_{21} |$.

\subsection{Muon anomalous magnetic moment}

The discrepancy between the current experimental result and the SM prediction is~\cite{Hagiwara:2011af,Keshavarzi:2018mgv,Bennett:2006fi,Davier:2010nc,Davier:2017zfy,Davier:2019can,Roberts:2010cj,Borsanyi:2020mff}
\begin{align}
\Delta a_{\mu} = a_{\mu}^{\rm exp} - a_{\mu}^{\rm SM} = 261(63)(48) \times 10^{-11}\,,
\label{eq:g-2exp}
\end{align}
where the numbers in the parentheses stand for the errors coming from $a_{\mu}^{\rm exp}$ and $a_{\mu}^{\rm SM}$, respectively. 
The current deviation is about $3.3\sigma$.~\footnote{Recently, new result from Fermilab has been reported~\cite{Abi:2021gix}, and the deviation from the SM prediction~\cite{Aoyama:2020ynm}, by combining with the previous result~\cite{Bennett:2006fi}, becomes $4.2\sigma$. } 
Therefore, if the anomaly is true, the new physics contribution should be positive to explain. 
However, it is well known that when there are only flavor diagonal couplings to ALP in the lepton sector, namely $(g^{\ell}_{L, R})_{ij} = 0$ ($i \neq j$), they can never explain the deviation by the one-loop contribution. 
Recently, it has been pointed out~\cite{Bauer:2019gfk} that ALPs can explain the deviation when we take into account the contribution from nonzero flavor off-diagonal elements, $(g^{\ell}_{L, R})_{ij} \neq 0$. 

As in Ref.~\cite{Bauer:2019gfk}, when we set $(g^{\ell}_A)_{22} / f = - 10^{-4} / {\rm TeV}$ and $c_{e \mu} / f \simeq 10 / {\rm TeV}$ with $m_a \simeq 0.12 \mathchar`- 0.15$ GeV, we find a parameter space to account for the deviation in $\Delta a_\mu$ without conflicting with several experimental bounds from lepton flavor violating processes. 
Furthermore, there arises also a contribution from the BZ type loop involving the $a \mathchar`-\gamma \mathchar`-\gamma$ coupling, $C_{\gamma \gamma}^{\rm eff}$ in Eq.~\eqref{a-gamma-gamma-app}. 
Therefore, we consider all these contributions and try to find the parameter space which explains the $(g-2)_{\mu}$ anomaly. 
Both two loop functions are available in Ref.~\cite{Bauer:2019gfk,Bauer:2017ris}.

\subsection{Charged lepton flavor violation}

The other relevant lepton-flavor violating process potentially induced by the ALPs is $\mu \to e \gamma$. 
This process is also related to $(g^{\ell}_A)_{22}$ in Eq.~\eqref{eq:Lagmass-app}, $c_{e \mu}$ in Eq.~\eqref{cemu}, and $C_{\gamma \gamma}^{\rm eff}$ in Eq.~\eqref{a-gamma-gamma-app}. 
The analytic formula for the decay width and the related loop function can be found in Ref.~\cite{Bauer:2019gfk}. 
The current experimental upper limit is~\cite{Tanabashi:2018oca} 
\begin{align}
{\rm BR}(\mu \to e \gamma) < 4.2 \times 10^{-13} \,\,\, \text{(90\% C.L.)}\,,
\label{eq:expmutoegam}
\end{align}
and the future prospect reported from MEG collaboration~\cite{Baldini:2013ke} (in three years of running) is 
\begin{align}
{\rm BR}(\mu \to e \gamma) < 6 \times 10^{-14}\,. 
\label{eq:expmutoegamftr}
\end{align}

\subsection{Limits on the ALP coupling to diphoton} 

In addition to flavor limits, the $a \mathchar`- \gamma \mathchar`-\gamma$ coupling $C_{\gamma \gamma}^{\rm eff}$ for the ALP mass around 140 MeV is bounded as seen from Fig.~\ref{fig:summary}: 
\begin{align}
\frac{0.0002}{\rm TeV} \lesssim \left| \frac{C_{\gamma \gamma}^{\rm eff}}{f} \right| \lesssim \frac{0.00164}{\rm TeV}, 
\,\,\, 
\frac{0.157}{\rm TeV} \lesssim \left| \frac{C_{\gamma \gamma}^{\rm eff}}{f} \right| \lesssim \frac{2297}{\rm TeV} \,.
\label{eq:Cggconst}
\end{align}
Note that the upper bounds in the right inequalities are the conservative limits from LEP searches~\cite{Acciarri:1995gy,Aaltonen:2013mfa,Jaeckel:2015jla,Bauer:2017ris} (see also the main text).

\end{document}